\documentstyle[multicol,aps,psfig,epsf]{revtex}
\newcommand{\beq}{\begin{equation}}
\newcommand{\eeq}{\end{equation}}
\newcommand{\bdis}{\begin{displaymath}}
\newcommand{\edis}{\end{displaymath}}
\newcommand{\bea}{\begin{eqnarray}}
\newcommand{\eea}{\end{eqnarray}}
\newcommand{\barr}{\begin{array}}
\newcommand{\earr}{\end{array}}
\newcommand{\lan}{\langle}
\newcommand{\ran}{\rangle}
\begin{document}
\twocolumn[\hsize\textwidth\columnwidth\hsize\csname 
 @twocolumnfalse\endcsname

\title{A ``Tetris''-like model for the Compaction of Dry Granular Media}
\author{Emanuele Caglioti$^{1}$, Vittorio Loreto$^{2,3}$, 
Hans J. Herrmann$^{2,4}$, and Mario Nicodemi$^{2,5}$}
\pagestyle{myheadings}
\markboth{Caglioti, Loreto, Herrmann and Nicodemi}{Caglioti, Loreto
  Herrmann and Nicodemi}
\address{
$^1$Dipartimento di Matematica, Universit\'a di 
Roma ``La Sapienza'', Piazzale Aldo Moro 2, 00185 Roma, 
Italy}
\address{
$^2$P.M.M.H. Ecole de Physique et Chimie Industrielles,
10, rue Vauquelin, 75231 Paris CEDEX 05 France}
\address{
$^3$ENEA Research Center, Localit\'a Granatello C.P. 32 80055 
Portici - Napoli, Italy}
\address{
$^4$ICA1, Univ. Stuttgart, Germany}
\address{
$^5$Dipartimento di Fisica, Universit\'a di Napoli ``Federico II'',
INFM and INFN Sezione di Napoli \\
Mostra d'Oltremare, Pad. 19, 80125 Napoli, Italy}

\maketitle
\date{\today}
\maketitle
\begin{abstract}
We propose a two-dimensional geometrical model, based on
the concept of geometrical frustration, conceived for the study of
compaction in granular media. 
The dynamics exhibits an interesting inverse logarithmic law 
that is well known from real experiments. 
Moreover we present a simple dynamical model of $N$ planes exchanging 
particles with excluded volume problems, which allows to clarify the origin of
the logarithmic relaxations and the stationary density distribution.
A simple mapping  allows us to cast this Tetris-like model in the form
of an Ising-like spin systems with vacancies.   
\end{abstract}
\smallskip
\vskip2pc]

A granular system may be in a number of different microscopic 
states at fixed macroscopic densities, and many unusual properties 
are linked to its non trivial packing \cite{JNScience92,Mehta}.
As pointed out by Edwards \cite{Mehta,Edwards,Hans} 
the role that the concept of free energy plays in standard thermal 
systems as Ising models, in granular media seems to be played by the 
``effective volume'', derived by a complex function
of grain positions and orientations. 
In this way statistical mechanics provides theoretical concepts in 
the context of non-thermal systems.

A recent experiment on the problem of density compaction 
in a dry granular system under tapping, has shown
\cite{Knight} that density compaction follows an inverse
logarithmic law with the tapping number.

Several approaches have been proposed to explain this behavior
\cite{BarkerMehta,Hong,Lintz,NCH,Ben-Naim,degennes}, as
geometrical models of ``parking'' \cite{Ben-Naim,evans} or
simple free-volume theories \cite{degennes} or
the study of the dynamics of a frustrated lattice gas with quenched
disorder subject to gravity and vibrations \cite{NCH}.
 
In many seemingly different cases 
the logarithmic relaxation proposed in \cite{Knight} to describe
experimental data is reproduced. Moreover the 
logarithmic law has turned out to be robust with respect to changes 
in the tapping procedure\cite{NCH}. 
This suggests that such a relaxation behavior is
extremely general and not linked to specific 
properties of definite realizations. 

Here we introduce a purely geometrical model of simple 
particles with several shapes on a lattice. We show that when 
subject to gravity and vibrations a logarithmic density relaxation
\cite{Knight} is found, due to the high entropic barriers 
(originated for geometrical reasons) to be passed
by particles to improve global packing.

\begin{figure}[h]
\centerline{
        \psfig{figure=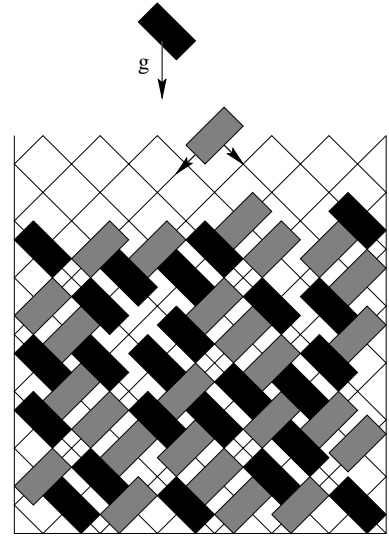,width=5cm,angle=-90}}

        \vspace*{0.5cm}
\caption{Schematic picture of one particular
configuration of the grains considered in the simplest version of the 
{\em Tetris}-like model.
The two types of particles have to fulfill only geometrical 
constraints in their dynamics. As shown in the figure these constraints
are due to the impossibility for the particles to overlap. }
\label{lattice}
\end{figure}

We imagine a model, similar to the computer game {\em Tetris} 
in which neighboring 
grains can find different packing volumes according to their 
relative geometrical orientations. 
Although one could imagine a rich variety of shapes and dimensions,
like in the real computer game, it is useful, without loss of
generality for the main features, to think just of a system of 
elongated particles which occupy the 
sites of square lattice tilted by $45^{\circ}$ (see Fig.~\ref{lattice}),  
with periodic boundary conditions in the horizontal direction
(cylindrical geometry) and a rigid plane at its bottom.
In general the only interactions between the particles are the
geometrical ones. Particles cannot overlap and this condition produces
very strong constraints (frustration) on their relative positions. For
instance in the simplest case of two kind of elongated
particles pointing in two (orthogonal) directions, the frustration
implies that two identical particles (pointing in the same direction)
cannot occupy neighboring sites in this direction.
The particles are in principle allowed to rotate if at least three of
their nearest neighbors are empty. This condition is such that 
for sufficiently high densities the rotation
events become negligible and the particles keep definitely their 
orientation. It is then reasonable, in the limit of a sufficiently 
large system, to consider an equal number of the two kind of
particles. There is no other form of 
interaction between particles, and in this sense the model is 
purely geometrical.

As stated, the particles are confined to a box and subjected to
gravity. The effect of vibrations is introduced by allowing the 
possibility of moving also upwards, as explained below.

The system is initialized by filling the container. The procedure of
filling consists in inserting the grains at the top of the system,
one at the time, and let them fall down, performing, under the 
effect of gravity, 
an oriented random walk on the lattice, until they reach 
a stable position, say a position in which they cannot fall further.
This filling procedure is realized by the addition of one
particle at the time and stops when no particles can enter the 
box from the top anymore.

In our case the dynamics can be divided in two alternating steps.  
First, in a {\em heating} process (tapping)
the system is perturbed by allowing the grains to move in any
allowed directions with a probability $p_{up}$ to move upwards 
(with $0 < p_{up} < 0.5$) and a  
probability $p_{down}=1-p_{up}$ to move downwards. 
After each tapping has been completed (i.e. a fixed number $N$  
 moves per particle have been attempted with a fixed value of 
$x=p_{up}/p_{down}$) we allow the system to relax setting $p_{up}=0$.
The relaxation process ({\em Cooling}) is supposed to be completed 
just when no particles can move anymore under just the effect of gravity, 
i.e. unless $p_{up}$ is switched on. 
After this relaxation the system is 
in a stable static state and one starts again the cycle.
We verified how the basic features of our model are very robust with
respect to variations in the Monte-Carlo procedure.
It is worth to stress how our dynamical procedure,
is very close physically to the real processes of vibro-compaction
\cite{vibro}. Work is in progress to implement in our system 
the method proposed in \cite{rosato}  which allows for the simulation
of a real tapping process.

More precisely the single dynamical step consists of the following 
operations: 1) extracting  with uniform probability a grain;
2) extracting a possible movement for this grain among 
the $4$ first neighbors ($2$ for the cooling process) according to
the probabilities $p_{up}$ and $p_{down}$; 3) move the grain if
{\bf all} the possible geometrical constraints with the neighbors are
satisfied.

We performed numerical simulations of the {\em Tetris}-like model 
in order to investigate its compaction properties.
In particular we measured the density of the packing, i.e. the 
percentage of sites occupied with respect to the total number of sites,
after each relaxation step and, in correspondence with real
experiments, we plot the behavior of this density as a function of 
the number of taps.
In order to avoid finite-size effects we considered systems with
a linear size of at least $L=50$ sites and, in order to be sure to observe
bulk effects, we measured the density in the lower $25\%$ of the system.

\begin{figure}[h]
\centerline{
        \psfig{figure=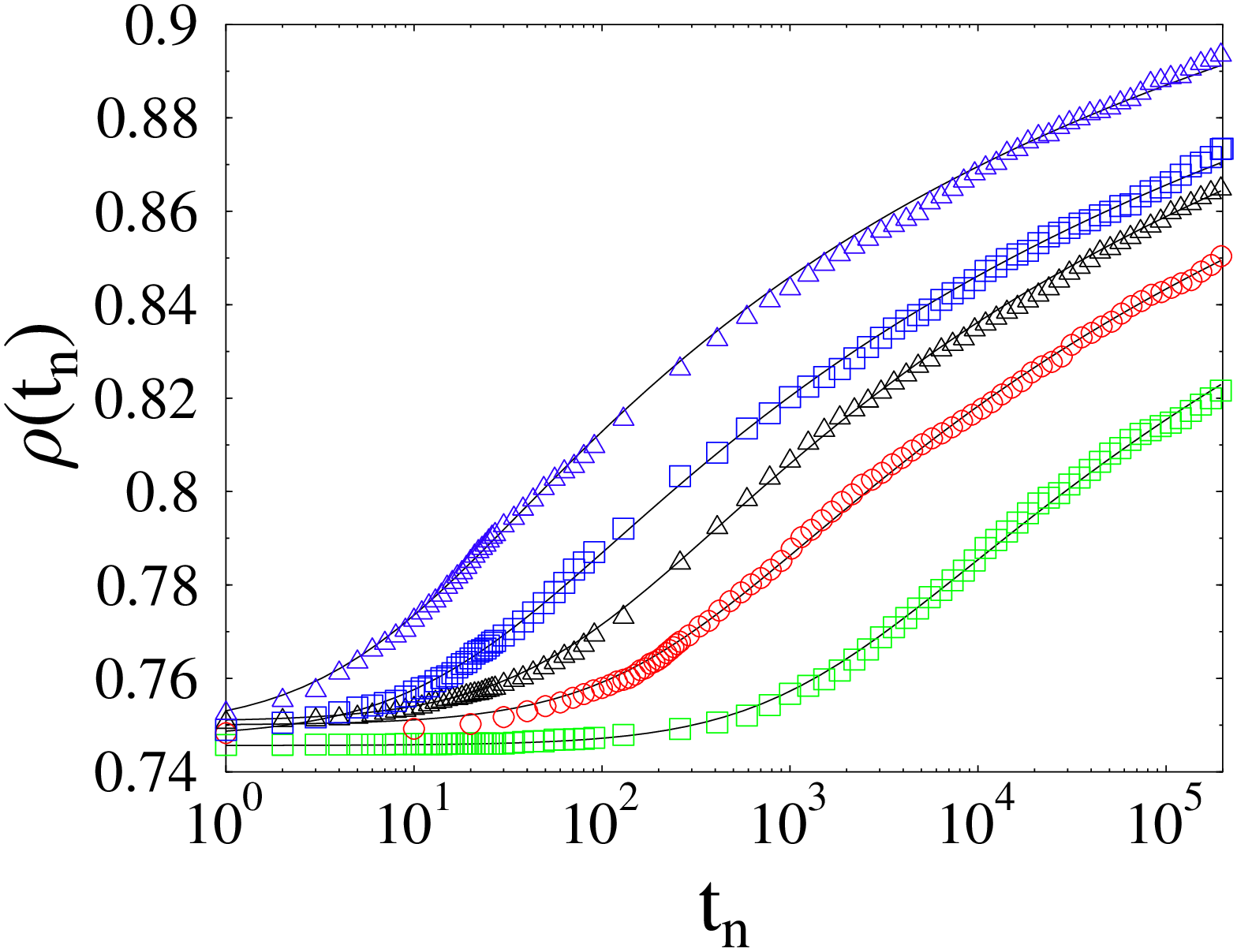,width=5cm,angle=-90}}
\caption{Logarithmic behavior of the density of the packing, measured in
  the lower $25\%$ of the system, as a function of tapping number
  $t_n$, for five different values of amplitude vibrations 
$x=p_{up}/p_{down}=0.001,0.01,0.03,0.1,0.5$, from bottom to top. 
The superimposed logarithmic fit curves, given by eq.(\ref{rho}), 
were proposed to describe experimental data.}
\label{d_log}
\end{figure}

Our main results on compaction 
are summarized in Fig.~\ref{d_log} which shows the
evolution of the density, as a function of the number of taps,
for different values of $x$ and for a system of dimension $L=50$.
The different curves, obtained with a tap length of one
iteration per particle, can be fitted according to the following
inverse logarithmic law:
\beq
\rho(t_{n})=\rho_{\infty} -\frac{\Delta
  \rho_{\infty}}{1+B \cdot \log(t_{n}/\tau+1)}
\label{rho}
\eeq
with $\rho_{\infty}=1$, a value $\Delta \rho_{\infty}=0.25$, 
which depends only on the loose packing density 
$\rho_{t_0}\sim 0.75$, and two free parameters,
$B$ and a characteristic time $\tau$, for which we observe an
algebraic dependence on $x$:
\begin{equation}
\tau = A x^{-\gamma} ~ .
\end{equation}
where $\gamma \simeq 0.84$ and $A=4.3$.
In this case $\tau$ has the meaning  of the minimum time over which
one starts to observe a compaction process. Up to times $t_{n} <<
\tau$, in fact, $\rho(t_n)$ keeps practically the initial value. A
complete and detailed analysis of these numerical results is reported
in \cite{nostro}.

Let us now briefly discuss how the system reaches the close-packing
density which, just in the case of the simplest version with
only two possible shapes, corresponds to a perfectly ordered state
with unitary density. It is worth to stress again how this
choice does not change the qualitative behavior of the system and an
infinity of disordered ground states can be obtained just allowing
for a rich variety of shapes for the particles\cite{nostro}.
The approach to this state, realized by means of the two-step
dynamics described above, represents a complex non-equilibrium
process in which the system evolves alternatively with two 
different ``temperatures'': a temperature $T_2$ (heating process) 
such that $e^{-{{2 g}\over {T_2}}}={p_{up} \over {1 - p_{up}}}$ and
a temperature $T_1=0$ for the cooling process. 
The first step could be considered as 
a process going towards equilibrium in which detailed balance holds.
Its features in many respects are very similar 
to the simple {\em hard-square model} \cite{GauFish}.
The step at zero temperature is an out of equilibrium process
which involves an irreversible positioning of the particles.
Globally the microscopic reversibility and detailed 
balance are lost.

In order to gain a deeper insight into the quoted logarithmic
dynamical behaviors, let us introduce and discuss a simple model 
which describes the evolution of a system of particles which hop 
on a lattice of $n=0,...,N$ stacked planes according to the ideas of 
``parking'' introduced in \cite{Ben-Naim,evans}.
We consider a system of particles which can move up 
or down between $N$ layers in such a way that their total number 
is conserved. We ignore the correlations among particles
rearrangements and the problem related to the mechanical stability of
the system. 
The master equation for the density on a generic plane $n$, except for
the $n=0$ plane, is given by:
\begin{eqnarray}
\partial_t \rho_n &=& 
(1-\rho_n)D(\rho_n) [ \rho_{n-1} \cdot p_{up} +\rho_{n+1} \cdot
p_{down}) ]+\nonumber \\
&& -\rho_n [(1-\rho_{n-1}) D(\rho_{n-1}) p_{down} +\\ 
&& +(1-\rho_{n+1}) D(\rho_{n+1})p_{up}]\nonumber  
\label{geneeq}
\end{eqnarray}
where $p_{down}$ and $p_{up}$ have been defined above for
the {\em Tetris}-like model. $D(\rho_{n})$ is a sort of mobility
for the particles given by the probability that the particle could
find enough space to move.  Apart form other effects it takes mainly 
onto account the geometrical effects of frustration, i.e. 
the fact that the packing prevents the free move of 
the particles.
In a naive way one could imagine a functional form like 
$D(\rho_{n})= \rho_n (1- \rho_{n^{\prime}})$ obtained by considering 
only the nearest neighbors interactions in the {\em Tetris}-like
model. It is easy to realize that such an approach
does not account for the complexity of the problem where the packing 
at high densities creates long range correlations in the system,
and, using this functional form, the equations show a trivial 
exponential relaxation.
Non trivial results are obtained with a careful choice of the
functional form for $D(\rho_{n})$ which takes onto account
the cooperative effects on the dynamics generated by the frustration. 
This functional form can be obtained by evaluating the rate specifying
how many steps are needed in a frustrated system, 
i.e. the {\em Tetris}-like model, with respect to a
non-frustrated one, to achieve a rearrangement in a new
configuration. We do not report here the complete
calculation of this rate which involves the evaluation of the 
configurational entropy of the frustrated and of the non-frustrated 
systems respectively, and we refer to \cite{nostro2}. 
It is nevertheless particularly enlightening
to consider the example $N=2$.
In this case the system can be reduced to a one-dimensional chain in which
the average length of the filled intervals turns out to be given by
$<l> \simeq \rho/(1-\rho)$. The number of steps to move a certain
particle, related to the number of steps necessary to move the
entire interval (of average length $<l>$) aside that particle,
will be the order of $N_{\rho} \sim exp<l>$.
The general form of $D(\rho_{n})$, although very complicate, must 
then include a term like
\beq
D(\rho_{n})=D_0\exp[-\rho_{n}/(1-\rho_{n})].
\label{drho}
\eeq
We checked this functional form, which can be seen how the
outcome of a free-volume theory for granular media
\cite{Ben-Naim,degennes,nostro2}, by
means of specific simulations for that quantity on the 
{\em Tetris}-like model\cite{nostro}.

In the general case with arbitrary $N$ we obtained 
the exact asymptotic stationary solution for the density on each 
plane. 
This solution is in an implicit form 
and for that we refer to\cite{nostro}. 
It is possible nevertheless to extract the approximate 
explicit behaviors. In particular, if $M$ is the total ``mass'' of 
the system, i.e. the maximal number of planes which can be 
completely filled, one gets:
\beq
\left\{
\begin{array}{l}
\rho_k^{\infty} \simeq 1 - 1/[(M-k) \cdot \log(1/x)] \,\, \mbox{for} \,\,
k << M\\\\
\rho_k^{\infty} \simeq e^{(M-k) \cdot \log(1/x)}\,\, \mbox{for} \,\,
k >> M.
\end{array}
\right.
\label{stationary}
\eeq
The stationary solution tends thus to a step function $\theta(k-M)$ in
the limit $x \rightarrow 0$.

Let us now comment on the dynamical behavior of the system, i.e.
the relaxation towards the stationary solution.
We start by considering the simplest case with just $N=2$ planes.
In the limit $x\equiv p_{up}/p_{down} <<1$
and for a sufficiently high total density, 
$\rho=1-\epsilon$ ($\epsilon<<1$), one can easily prove
\cite{nostro} that the asymptotic equilibrium lower plane
density behaves like $\rho_1^{\infty}\simeq 1+1/\log[xf(\epsilon)]$ where  
$f(\epsilon)=2 \epsilon/(1-2 \epsilon)\exp[-(1-2\epsilon)/
{2 \epsilon}]$. 
The dynamical equation for $\rho_1$ can be written exactly.
In the limit $\rho_1 >> (1- 2 \epsilon)$ (which holds for sufficiently
long times) this equation exhibits a very simple form as:
\beq
\partial_t \rho_1 = B(x,\epsilon)
(1-\rho_1)D(\rho_1)-A(x,\epsilon)\rho_1
\label{appE}
\eeq
with $B(x,\epsilon)=(1-2 \epsilon)/(1+x)$ and 
$A(x,\epsilon)=2 \epsilon\exp[-(1-2 \epsilon)/
{2 \epsilon}]x/(1+x)$. 
This equation has the same form of the one dimensional 
``parking problem'' studied in \cite{Ben-Naim} whose absorption 
and desorption parameters are now written in terms of the global 
density of the system, $1-\epsilon$, and of the vibration 
amplitude ratio $x$. It exhibits a
logarithmic solution up to times of the order of $t_0\sim
1/A(x,\epsilon)$ \cite{Ben-Naim}. Later on, when the
density approaches its steady state value, the first term on the
right-hand side of eq.(\ref{appE}) becomes negligible with respect
to the loss term and an exponential saturation becomes dominant. 
Here we just note that 
$t_0$ grows inversely proportional to $x$ but has an essential
singularity for $\epsilon$ going to zero. So for low enough
amplitude vibrations or high enough densities the logarithmic region
extends actually up to any experimentally observable time.
The cooperative effect of interaction among the different planes 
makes the times over which one observes the logarithmic relaxation 
longer and longer. 
Crucial for this effect is the value of $\epsilon$,
i.e. the asymptotic difference of density between two adjacent planes.  
In the general case of $N$ planes, one has from eq.(\ref{stationary})
that in the bulk, for finite values of $x$, $\epsilon_k \simeq
x^{M-k}$, i.e. the $\epsilon_k$ are exponentially small in $M$.
We then expect that the logarithmic relaxations extends up to times
of the order of $x^{-M}$ (see \cite{nostro} for a detailed discussion
of this point).


Let us now notice a further aspect of our model.
The two-step dynamics of our model may be easily interpreted in terms 
of a Glauber dynamics for an Hamiltonian with Ising-like variables.
In this language the geometrical model
is mapped into the following Ising-like Hamiltonian with vacancies
in the limit $J\rightarrow\infty$: 
\begin{equation}
H=\sum_{\lan ij\ran } J(S_i S_j -a_{ij}(S_i+S_j)-1)n_i n_j+g \sum_i y(i)
\label{H}
\end{equation}
where $n_i=0,1$ are occupancy variables and $S_i=\pm 1$ are spin
variables that corresponds to the twofold orientation of the particles. 
$a_{ij}=\pm 1$ are fixed non-random bond fields with an ordered
structures: $a_{ij}= 1$ for bonds along one direction of the lattice
and $a_{ij}= -1$ for bonds in the other.
In the gravitational term, $-g \sum_i y(i)$, $g$ is the gravity, 
and $y(i)$ is the ordinate of the lattice
site $i$. It is easy to realize that the 
sum of the $a_{ij}$ converging on each single site is zero. 
This implies that the ground
state of Hamiltonian~(\ref{H}) is perfectly antiferromagnetic if the
densities of the two kinds of particles are equal.
This state is reached just when all the sites of the lattice
are occupied, so $n_i=1 \,\, \forall i$.
This mapping, and the ones for the models with a variety of shapes
which lead to Potts-like Hamiltonians, are particularly useful as
starting points for an analysis of these systems in a thermodynamic
framework \cite{nostro}.


In this paper we have introduced a very simple geometrical
model in order to describe the phenomenon of compaction in dry
granular media. It takes in account excluded volume
effects, say the geometrical constraints which are felt by granular 
media during the relaxation towards the highest density optimal packing
configuration. When subjected to Monte Carlo vibrations, defined by 
a diffusive dynamics, it exhibits a density compaction after tapping 
which reproduces the inverse logarithmic behavior found in both 
experiments \cite{Knight} and other models \cite{Lintz,NCH,Ben-Naim}. 

This {\em Tetris}-like model can be easily generalized by introducing 
an arbitrary fixed number of shapes for the particles which correspond to 
complicated matrices for the particle-particle interactions.
This kind of generalization does not change the
qualitative structure of the relaxation but it could account
for other effects of disorder in granular media: segregation,
hysteresis etc. \cite{nostro}.

Furthermore we presented a simple dynamical model of $N$ planes 
exchanging particles with excluded volume effects. 
For this model we have found the exact stationary density 
distribution and we have shown how, without loss of generality with 
respect to the choice of particular geometrical constraints, it 
allows for an explanation of the inverse logarithmic law for 
compaction. 

It is moreover interesting that the pure geometrical model 
presented here can be mapped into a simple Hamiltonian formalism 
of an Ising antiferromagnet. This connects our work to
previous works \cite{Edwards,NCH,Ben-Naim} introduced to
discuss different aspects of granular media phenomenology, 
and could open the way to their systematic analysis \cite{nostro}.
{\bf \large Acknowledgments} We are indebted with A. Coniglio for
useful suggestions. We thank P.G. De Gennes for bringing to our
attention his preprint.

\end{document}